\documentstyle[11pt]{article}
\newcommand{\be}{\begin{equation}}
\newcommand{\ee}{\end{equation}}
\begin{document}
\title{The circle electromagnetic pulsar}
\author{Miroslav Pardy\\
Department of Physical Electronics \\
Masaryk University, Faculty of Science \\
Kotl\'a\v rsk\'a 2, 611 37 Brno, Czech
Republic, \\
e-mail: pamir@physics.muni.cz}
\date{\today}
\maketitle

\large

\begin{abstract}
The power spectrum formula of the synchrotron radiation generated by the
electron and positron moving along the concentric circles
at the opposite angular
velocities in homogenous magnetic field
is derived in the Schwinger version of quantum electrodynamics.
The asymptotical form of this formula is found.
The spectrum
depends periodically on time which means
that the system composed from electron, positron and magnetic
field forms the pulsating system. The similar calculations are performed
in the case that the circles of motion are excentric.
\end{abstract}


\baselineskip 17pt

\section{Introduction}

   The production of photons by circular motion of charged particle
in accelerator is one of the most interesting problems in the classical
and quantum electrodynamics.

In this paper we are interested in the synergic photon production
initiated by the circular motion of electron and positron
in the homogenous magnetic field.
It is supposed that electron and positron are moving at
the opposite angular velocities.
This process is the generalization of the one-charge synergic
synchrotron-\v{C}erenkov
radiation which has been calculated in source theory two decades ago
by Schwinger et al. [1]. We will follow also the article [2] as
the starting point.
Although our final problem is the radiation of the two-charge system in vacuum,
we consider, first in general, the presence of dielectric medium, which is
represented by the
phenomenological index of refraction $n$.
Introducing the phenomenological constant enables to consider
also the \v{C}erenkovian processes.

We will investigate here how the original Schwinger et al. spectral
formula of the synergic synchrotron-\v{C}erenkov radiation of the
charged particle is modified if the electron and positron are
moving along the concentric and excentric circles at
the opposite  angular velocities.
This problem is an analogue of the linear problem solved recently
by author [3] also in source theory.

We apply the Schwinger methods [4--6] which was initially used
in a description of the particle physics situations occurring in
high-energy physics experiments. It enables
simplification of the calculations in the
electrodynamics and gravity where the interactions are mediated by
the photon or graviton, respectively. It simplifies particularly the
calculations with radiative corrections [6,7].

\section{Formulation of a problem}

The basic formula of the Schwinger source theory is the so called
vacuum to vacuum amplitude:
\be
\label{1}
\langle 0_{+}|0_{-} \rangle = e^{\frac{i}{\hbar}\*W},
\ee
where in case of the electromagnetic field in the medium, the action
$W$ is given by the following formula:
\be
\label{2}
W = \frac{1}{2c^2}\*\int\,(dx)(dx')J^{\mu}(x){D}_{+\mu\nu}(x-x')J^{\nu}(x'),
\ee
where
\be
\label{3}
{D}_{+}^{\mu\nu} = \frac{\mu}{c}[g^{\mu\nu} +
(1-n^{-2})\beta^{\mu}\beta^{\nu}]\*{D}_{+}(x-x'),
\ee
where $\beta^{\mu}\, \equiv \, (1,{\bf 0})$, $J^{\mu}\, \equiv \,(c\varrho,{\bf
J})$ is the conserved current, $\mu$ is the magnetic permeability of
the medium, $\varepsilon$ is the dielectric constant od the medium and
$n=\sqrt{\varepsilon\mu}$ is the index of refraction of the medium.
Function ${D}_{+}$ is defined as follows [1]:

\be
\label{4}
D_{+}(x-x') =\frac {i}{4\pi^2\*c}\*\int_{0}^{\infty}d\omega
\frac {\sin\frac{n\omega}{c}|{\bf x}-{\bf x}'|}{|{\bf x} - {\bf x}'|}\*
e^{-i\omega|t-t'|}.
\ee

The probability of the persistence of vacuum follows from the vacuum
amplitude (1) in the following form:

\be
\label{5}
|\langle 0_{+}|0_{-} \rangle|^2 = e^{-\frac{2}{\hbar}\*\rm Im\*W},
\ee
where ${\rm Im}\;W$ is the basis for the definition of the spectral
function $P(\omega,t)$ as follows:
\be
\label{6}
-\frac{2}{\hbar}\*{\rm Im}\;W \;\stackrel{d}{=} \; -\,
\int\,dtd\omega\frac{P(\omega,t)}{\hbar\omega}.
\ee

Now, if we insert eq. (2) into eq. (6), we get
after extracting $P(\omega,t)$ the following general expression
for this spectral function:

$$P(\omega,t) = -\frac{\omega}{4\pi^2}\*\frac{\mu}{n^2}\*\int\,d{\bf x}
d{\bf x}'dt'\left[\frac{\sin\frac{n\omega}{c}|{\bf x} -
{\bf x}'|}{|{\bf x} - {\bf x}'|}\right]\;\times $$

\be
\label{7}
\cos[\omega\*(t-t')]\*[\varrho({\bf x},t)\varrho({\bf x}',t')
- \frac{n^2}{c^2}\*{\bf J}({\bf x},t)\cdot{\bf J}({\bf x}',t')].
\ee

Let us recall that the last formula can be derived also in the classical
electrodynamical context as it is shown for instance in the
Schwinger article [8].
The derivation of the power spectral formula from the
vacuum amplitude is more simple.

\section{The power spectral formula of motion of opposite charges}

Now, we will apply the formula (7) to the two-body system
with the opposite charges moving at the oposite angular velocities
in order to get in general synergic synchrotron-\v Cerenkov
radiation of electron and positron moving in a uniform
magnetic field.

While the synchrotron radiation is generated in a vacuum, the synergic
synchrotron-\v Cerenkov radiation can produced only in a
medium with dielectric constant $n$.
We suppose the circular motion with velocity ${\bf v}$
in the plane perpendicular to the
direction of the constant magnetic field ${\bf H}$ (chosen to be
in the $+z$ direction).

We can write the following formulas for the
charge density $\varrho$ and for the current
density ${\bf J}$ of the two-body system with opposite charges and opposite
angular velocities:

\be
\label{8}
\varrho({\bf x},t) = e\*\delta\*({\bf x}-{\bf x_{1}}(t))
-e\*\delta\*({\bf x}-{\bf x_{2}}(t))
\ee
and

\be
\label{9}
{\bf J}({\bf x},t) = e\*{\bf v}_{1}(t)\*\delta\*({\bf x}-{\bf x_{1}}(t))
-e\*{\bf v}_{2}(t)\*\delta\*({\bf x}-{\bf x_{2}}(t))
\ee
with
\be
\label{10}
{\bf x}_{1}(t)  = {\bf x}(t) =
R({\bf i}\cos(\omega_{0}t) + {\bf j}\sin(\omega_{0}t)),
\ee

\be
\label{11}
{\bf x}_{2}(t) =
R({\bf i}\cos(-\omega_{0}t) +
{\bf j}\sin(-\omega_{0}t) =
{\bf x}(-\omega_{0},t) = {\bf x}(-t).
\ee

The absolute values of velocities of both particles are
the same, or $|{\bf v}_{1}(t)| =  |{\bf v}_{2}(t)| = v$, where ($H = |{\bf H}|, E =$ energy of a particle)
\be
\label{12}
{\bf v}(t) = d{\bf x}/dt, \hspace{5mm} \omega_{0} = v/R, \hspace{5mm}
R = \frac {\beta\*E}{eH}, \hspace{5mm}
\beta = v/c, \hspace{5mm} v = |{\bf v}|.
\ee

After insertion of eqs. (8)--(9) into eq. (7), and after some mathematical
operations we get

$$P(\omega,t) =
-\frac{\omega}{4\pi^2}\*\frac{\mu}{n^2}e^{2}\*\int_{-\infty}^{\infty}\,
dt'\cos(t-t')\sum_{i,j = 1}^{2}(-1)^{i+j}
\left[1 - \frac {{\bf v}_{i}(t)\cdot {\bf v}_{j}(t')}{c^{2}}n^{2}\right]
\;\times $$

\be
\label{13}
\left\{\frac{\sin\frac {n\omega}{c}|{\bf x}_{i}(t) -{\bf x}_{j}(t')|}
{|{\bf x}_{i}(t) -{\bf x}_{j}(t')|}\right\}.
\ee

Using $t' = t + \tau$, we get for

\be
\label{14}
{\bf x}_{i}(t) -{\bf x}_{j}(t')
\stackrel{d}{=} {\bf A}_{ij},
\ee

\be
\label{15}
|{\bf A}_{ij}| = [R^{2} + R^{2} - 2RR\cos(\omega_{0}\tau +
\alpha_{ij})]^{1/2} =
2R\left|\sin\left(\frac {\omega_{0}\tau + \alpha_{ij}}{2}\right)\right|,
\ee
where $\alpha_{ij}$ were evaluated as follows:
\be
\label{16}
\alpha_{11} = 0,\quad \alpha_{12 } = 2\omega_{0}t,
\quad \alpha_{21} = 2\omega_{0}t, \quad  \alpha_{22} = 0.
\ee
Using

\be
\label{17}
{\bf v}_{i}(t)\cdot{}{\bf v}_{j}(t+\tau) = \omega_{0}^{2}R^{2}
\cos(\omega_{0}\tau +
\alpha_{ij}),
\ee
and relation (15) we get with $v= \omega_{0}R$

$$P(\omega,t) =
-\frac{\omega}{4\pi^2}\*\frac{\mu}{n^2}e^{2}\*\int_{-\infty}^{\infty}\,
d\tau \cos\omega\tau \sum_{i,j = 1}^{2}(-1)^{i+j}
\left[1 - \frac {n^{2}}{c^{2}}v^{2}\cos(\omega_{0}\tau + \alpha_{ij})\right]
\;\times $$

\be
\label{18}
\left\{\frac{\sin\left[\frac {2Rn\omega}{c}
\sin\left(\frac {(\omega_{0}\tau + \alpha_{ij})}
{2}\right)\right]}
{2R\sin\left(\frac {(\omega_{0}\tau + \alpha_{ij})}{2}\right)}\right\}.
\ee

Introducing new variable $T$ by relation

\be
\label{19}
\omega_{0}\tau + \alpha_{ij} = \omega_{0}T
\ee
for every integral in eq. (18),
we get $P(\omega,t)$ in the following form

$$P(\omega,t) =
-\frac{\omega}{4\pi^2}\frac {e^{2}}{2R}
\*\frac{\mu}{n^2}\*\int_{-\infty}^{\infty} dT \sum_{i,j=1}^{2}(-1)^{i+j}
\times $$

\be
\label{20}
\cos(\omega T - \frac {\omega}{\omega_{0}}\alpha_{ij})
\left[1 - \frac {c^{2}}{n^{2}}v^{2}\cos(\omega_{0} T) \right]
\left\{\frac{\sin\left[\frac {2Rn\omega}{c}\sin
\left(\frac {\omega_{0}T}{2}\right)\right]}
{\sin\left(\frac {\omega_{0}T}{2}\right)}\right\}.
\ee
The last formula can be written in the more compact form,

\be
\label{21}
P(\omega,t) = -\frac {\omega}{4\pi^{2}}\frac {\mu}{n^{2}}\frac {e^{2}}{2R}
\sum_{i,j=1}^{2}(-1)^{i+j}\left\{P_{1}^{(ij)} -\frac {n^{2}}{c^{2}}v^{2}
P_{2}^{(ij)}\right\},
\ee
where

\be
\label{22}
P_{1}^{(ij)} = J_{1a}^{(ij)}\cos\frac {\omega}{\omega_{0}}\alpha_{ij} +
J_{1b}^{(ij)}\sin\frac {\omega}{\omega_{0}}\alpha_{ij}
\ee
and

\be
\label{23}
P_{2}^{(ij)} = J_{2A}^{(ij)}
\cos\frac {\omega}{\omega_{0}}\alpha_{ij} +
J_{2B}^{(ij)}\sin\frac {\omega}{\omega_{0}}\alpha_{ij},
\ee
where

\be
\label{24}
J_{1a}^{(ij)} = \int_{-\infty}^{\infty}dT\cos\omega T
\left\{\frac{\sin\left[\frac {2Rn\omega}{c}\sin
\left(\frac {\omega_{0}T}{2}\right)\right]}
{\sin\left(\frac {\omega_{0}T}{2}\right)}\right\},
\ee

\be
\label{25}
J_{1b}^{(ij)} = \int_{-\infty}^{\infty}dT\sin\omega T
\left\{\frac{\sin\left[\frac {2Rn\omega}{c}\sin
\left(\frac {\omega_{0}T}{2}\right)\right]}
{\sin\left(\frac {\omega_{0}T}{2}\right)}\right\},
\ee

\be
\label{26}
J_{2A}^{(ij)} = \int_{-\infty}^{\infty}dT\cos\omega_{0}T\cos\omega T
\left\{\frac{\sin\left[\frac {2Rn\omega}{c}\sin
\left(\frac {\omega_{0}T}{2}\right)\right]}
{\sin\left(\frac {\omega_{0}T}{2}\right)}\right\},
\ee

\be
\label{27}
J_{2B}^{(ij)} = \int_{-\infty}^{\infty}dT\cos\omega_{0}T\sin\omega T
\left\{\frac{\sin\left[\frac {2Rn\omega}{c}\sin
\left(\frac {\omega_{0}T}{2}\right)\right]}
{\sin\left(\frac {\omega_{0}T}{2}\right)}\right\},
\ee

Using

\be
\label{28}
\omega_{0}T = \varphi + 2\pi\*l,  \hspace{7mm} \varphi\in(-\pi,\pi),\;
\quad l = 0,\, \pm1,\, \pm2,\, ... ,
\ee
we can transform the $T$-integral into the sum of the telescopic
integrals according to the scheme:

\be
\label{29}
\int_{-\infty}^{\infty}dT\quad\longrightarrow \quad\frac {1}{\omega_{0}}
\sum_{l = -\infty}^{l = \infty}\int_{-\pi}^{\pi}d\varphi.
\ee

Using the fact that for the odd functions $f(\varphi)$ and $g(l)$,
the relations are valid

\be
\label{30}
\int_{-\pi}^{\pi}f(\varphi)d\varphi = 0; \quad \sum_{l=-\infty}^{l =
\infty}g(l) = 0,
\ee
we can write

\be
\label{31}
J_{1a}^{(ij)} = \frac {1}{\omega_{0}}\sum_{l}\int_{-\pi}^{\pi}
d\varphi\left\{\cos{\frac {\omega}{\omega_{0}}\varphi\cos{2\pi l}
\frac{\omega}{\omega_{0}}}\right\}
\left\{\frac{\sin\left[\frac {2Rn\omega}{c}\sin
\left(\frac {\varphi}{2}\right)\right]}
{\sin\left(\frac {\varphi}{2}\right)}\right\},
\ee

\be
\label{32}
J_{1b}^{(ij)} = 0.
\ee

For integrals with indices A, B we get:

\be
\label{33}
J_{2A}^{(ij)} = \frac {1}{\omega_{0}}\sum_{l}\int_{-\pi}^{\pi}
d\varphi\cos\varphi
\left\{\cos{\frac {\omega}{\omega_{0}}\varphi\cos{2\pi l}
\frac{\omega}{\omega_{0}}}\right\}
\left\{\frac{\sin\left[\frac {2Rn\omega}{c}\sin
\left(\frac {\varphi}{2}\right)\right]}
{\sin\left(\frac {\varphi}{2}\right)}\right\},
\ee

\be
\label{34}
J_{2B}^{(ij)} = 0,
\ee

So, the power spectral formula (21) is of the form:

\be
\label{35}
P(\omega,t) = -\frac {\omega}{4\pi^{2}}\frac {\mu}{n^{2}}\frac {e^{2}}{2R}
\sum_{i,j=1}^{2}(-1)^{i+j}\left\{P_{1}^{(ij)} - n^{2}\beta^{2}
P_{2}^{(ij)}\right\};\quad \beta = \frac {v}{c},
\ee
where
\be
\label{36}
P_{1}^{(ij)} = J_{1a}^{(ij)}\cos\frac {\omega}{\omega_{0}}\alpha_{ij}
\ee

\be
\label{37}
P_{2}^{(ij)} = J_{2A}^{(ij)}\cos\frac {\omega}{\omega_{0}}\alpha_{ij}.
\ee
Using the Poisson theorem

\be
\label{38}
\sum_{l = -\infty}^{\infty}\cos 2\pi\frac {\omega}{\omega_{0}}l  =
\sum_{k=-\infty}^{\infty}\omega_{0}\delta(\omega - \omega_{0}l),
\ee
we get for $J_{1a}^{(ij)}$ and $J_{2A}^{(ij)}$ ($z = 2ln\beta$):

\be
\label{39}
J_{1a}^{(ij)} = \sum_{l}\int_{-\pi}^{\pi}
d\varphi \cos\varphi\cos l\varphi
\left\{\frac{\sin(z\sin(\varphi/2))}
{\sin(\varphi/2)}\right\},
\ee

\be
\label{40}
J_{2A}^{(ij)} = \sum_{l}\int_{-\pi}^{\pi}
d\varphi\cos\varphi\sin l\varphi
\left\{\frac{\sin(z\sin(\varphi/2))}
{\sin(\varphi/2)}\right\},
\ee

Using the definition of the Bessel functions $J_{2l}$ and their
corresponding derivation and integral

\be
\label{41}
\frac {1}{2\pi}\int_{-\pi}^{\pi}d\varphi\cos\left(z\sin\frac {\varphi}{2}
\right)\cos l\varphi  = J_{2l}(z),
\ee

\be
\label{42}
\frac {1}{2\pi}\int_{-\pi}^{\pi}d\varphi\sin\left(z\sin\frac {\varphi}{2}
\right)\cos l\varphi  = - J'_{2l}(z),
\ee

\be
\label{43}
\frac {1}{2\pi}\int_{-\pi}^{\pi}d\varphi
\frac{\sin\left(z\sin\frac{\varphi}{2}\right)}
{\sin(\varphi/2)}\cos l\varphi  = \int_{0}^{z}J_{2l}(x)dx,
\ee
and using  equations

\be
\label{44}
\sum_{i,j = 1}^{2}(-1)^{i+j}\cos\frac {\omega}{\omega_{0}}\alpha_{ij}
= 2(1-\cos 2\omega t) = 4\sin^{2}\omega t,
\ee

and the definition of the partial power spectrum $P_{l}$

\be
\label{45}
P(\omega) = \sum_{l=1}^{\infty}  \delta(\omega - l\omega_{0})P_{l},
\ee
we get the following final form of the partial power spectrum
generated by motion of
two-charge system moving in the cyclotron:

\be
\label{46}
P_{l}(\omega,t) = [4(\sin\omega t)^{2}]
\frac {e^2}{\pi\*n^2}\*\frac {\omega\mu\omega_{0}}{v}\*
\left(2n^2\beta^2J'_{2l}(2ln\beta) -
(1 - n^2\*\beta^2)\*\int_{0}^{2ln\beta}dxJ_{2l}(x)\right).
\ee

So we see that the spectrum generated by the system of electron and positron
is formed in such a way that the original synchrotron spectrum generated by
electron is modulated by function
$4\sin^{2}(\omega t)$. This formula is analogical to the formula
derived in [3] for the linear motion of the two-charge
system emitting the \v Cerenkov radiation. The derived formula
involves also the synergic process composed from
the synchrotron radiation and the
\v Cerenkov radiation for electron velocity $v > c/n$ in a medium.

Our  goal is to apply the last formula
in situation where there  is  a vacuum. In
this case we can put $\mu = 1, n = 1$ in the last formula and so we have

\be
\label{47}
P_{l}(\omega,t) =
4 \sin^{2}\left(\omega t\right)
\frac {e^2}{\pi}\*\frac {\omega\omega_{0}}{v}\*
\left(2\beta^2J'_{2l}(2l\beta) -
(1 - \beta^2)\*\int_{0}^{2l\beta}dxJ_{2l}(x)\right).
\ee

So, we see, that final formula describing the
opposite motion of eletron and positron in accelerator is
of the form

\be
\label{48}
P_{l}(\omega,t) =
4 \sin^{2}\left(\omega t\right)P_{l(electron)}\left(\omega\right),
\ee
where $P_{electron}$ is the spectrum of radiation only of electron.
The result is suprising because we naively
expected that
the total radiation of the opposite charges should be

\be
\label{49}
P_{l}(\omega,t) =
P_{l(electron)}\left(\omega, t \right) +
P_{l(positron)}\left(\omega, t \right).
\ee

So, we see that the resulting radiation  can not be considerred
as generated by the isolated particles but by a synergical production of
a system of particles and magnetic field. At the same time we cannot
interprete the result as a result of interference of two sources
because the distance between sources radically changes and so, the
condition of an interference is not fulfilled.

The classical electrodynamics is not broken by this formula but what is broken
is our naive image on the processes in the magnetic
field. From the last formula also follows that at time $t = \pi k/\omega $
there is no radiation of the frequency $\omega$.
At every frequency $\omega$ the spectrum oscillates with frequency $\omega$.
If the radiation were generated not synergically, then
the spectral formula would be composed from two parts corresponding
to two isolated sources.

\section{The more realistic situation}

The situation which we have analysed was the ideal situation where the angle of collision of positron and electron
was equal to $\pi$. Now, the question arises what is the modification of a spectral formula when the collision angle between
particles differs from $\pi$. The goal of this article is to formulate this problem using the above mathematical approach.
It can be easily seen that if the second particle follows the shifted circle trajectory, than the collision angle
differs from $\pi$.  Let us suppose that the the circle center of the second particle
has coordinates (a,0). It can be easy to
see from the geometry of the situation and from the planimetry
that the collision angle is $ \pi - \alpha, \alpha
 \approx \tan \alpha \approx a/R$ where
$R$ is a radius of the first or second circle. The same result follows
from the analytical geometry of the situation.

While the equation of the first particle is the equation of the original trajectory, or this is eq. (10)

\be
\label{50}
{\bf x}_{1}(t) = {\bf x}(t) = R({\bf i}\cos(\omega_0 t) + {\bf j}\sin(\omega_0 t)),
\ee

the equation of a circle with a shifted center is as follows

\be
\label{51}
{\bf x}_{2}(t) = {\bf x}(t) = R({\bf i}(\frac {a}{R}+\cos(-\omega_0 t)) +
{\bf j}\sin(-\omega_0 t)) = {\bf x(-t)} + {\bf i}a,
\ee

The absolute values of velocities of both particles are equal and the
relation (12) is valid.
Instead of equation (14) we have for radius vectors of particle trajectories:

\be
\label{52}
{\bf x}_{i}(t) -{\bf x}_{j}(t')
\stackrel{d}{=} {\bf B}_{ij},
\ee
where ${\bf B}_{11} = {\bf A}_{11}, {\bf B}_{12} = {\bf A}_{12} - {\bf i}a, {\bf B}_{21} = {\bf A}_{21} + {\bf i}a,
{\bf B}_{22} = {\bf A}_{22}$.

In general, we can write the last information on
coeficients ${\bf B}_{ij}$ as follows:

\be
\label{53}
 {\bf B}_{ij} = {\bf A}_{ij} + \varepsilon_{ij}{\bf i}a ,
\ee
where $\varepsilon_{11} = 0, \varepsilon_{12} = -1, \varepsilon_{21} = 1,
\varepsilon_{22} = 0.$

For motion of particles along trajectories the absolute value of vector
${\bf A}_{ij} \gg a$ during the most part of the trajectory.
It means, we can determine
$B_{ij}$ approximatively. After elementary operations, we get:

\be
\label{54}
|{\bf B_{ij}}| =  (A_{ij}^{2} +
2|{\bf A}_{ij}|\varepsilon_{ij}a\cos\varphi_{ij} +
a^{2}\varepsilon_{ij}^{2})^{1/2},
\ee
where $\cos\varphi_{ij}$ can be expressed by the $x$-component of
vector ${\bf A}_{ij}$ and $|{\bf A}_{ij}|$ as follows:

\be
\label{55}
\cos\varphi_{ij} =  \frac{(A_{ij})_{x}}{|{\bf A}_{ij}|}
\ee

After elemetary trigonometric operations, we derive the following
formula  $(A_{ij})_{x}$.

\be
\label{56}
(A_{ij})_{x} = 2R\sin\frac {2\omega_{0}t + \omega_{0}\tau}{2}
\sin\frac {\omega_{0}\tau}{2}
\ee

Then, using equation (56), we get with $\varepsilon = a/R$.

\be
\label{57}
|{\bf B_{ij}}| = 2R (\sin^{2}\frac {\omega_{0}\tau + \alpha_{ij}}{2}
+ \varepsilon\varepsilon_{ij}\sin\frac {2\omega_{0}t + \omega_{0}\tau}{2}
\sin\frac {\omega_{0}\tau}{2} +
\varepsilon^{2}\frac{\varepsilon_{ij}^{2}}{4})^{1/2},
\ee

In order to perform the $\tau$-integration the substitution must be
introduced. However, the substitution $\omega_{0}\tau + \alpha_{ij} =
\omega_{0}T$ does not work. So we define the substitution $\tau = \tau(T)$
by the following transcendental equation (we will neglect the term with
$\varepsilon^{2}$):

\be
\label{58}
\left[\sin^{2}\frac {\omega_{0}\tau + \alpha_{ij}}{2}
+ \varepsilon\varepsilon_{ij}\sin\frac {\omega_{0}t + \omega_{0}\tau}{2}
\sin\frac {\omega_{0}\tau}{2}\right]^{1/2} =  \sin\frac {\omega_{0}T}{2}
\ee

Or, after some trigonometrical modifications  and using
the approximative formula $(1 + x)^{1/2} \approx 1 + x/2$ for $x\ll 1$

\be
\label{59}
\left[ ./. \right]^{1/2} \approx
\sin \left(\frac {\omega_{0}\tau + \alpha_{ij}}{2}\right)
+ \frac{\varepsilon}{2}\varepsilon_{ij}
\sin\left(\frac {2\omega_{0}\tau + 2\omega_{0}t -\alpha_{ij}}{2}\right)
=  \sin\frac {\omega_{0}T}{2}
\ee

We se that for $\varepsilon = 0$ the substitution is
$\omega_{0}\tau + \alpha_{ij} = \omega_{0}T$. The equation (58) is the
transcendental equation and the exact solution is the function
$\tau = \tau(T)$. We are looking for the solution of equation (59) in the
approximative form using the approximation $\sin x \approx x$.

Then, instead of (59) we have:

\be
\label{60}
 \left(\frac {\omega_{0}\tau + \alpha_{ij}}{2}\right)
+ \frac{\varepsilon}{2}\varepsilon_{ij}
\left(\frac {2\omega_{0}\tau + 2\omega_{0}t -\alpha_{ij}}{2}\right)
= \frac {\omega_{0}T}{2}
\ee
Using substitution

\be
\label{61}
\omega_{0}\tau + \alpha_{ij} = \omega_{0}T + \omega_{0}\varepsilon A
\ee
in eq. (60) we get, to the first order in $\varepsilon$-term:

\be
\label{62}
A = -\frac{\varepsilon_{ij}}{2\omega_{0}}(\omega_{0}T - 2\alpha_{ij} +
2\omega_{0}t)
\ee

Then, after some algebraic manipulation we get:

\be
\label{63}
\omega_{0}\tau  +  \alpha_{ij} =
\omega_{0}T(1 -\frac{\varepsilon}{2} \varepsilon_{ij})
- \varepsilon\varepsilon_{ij}\omega_{0}t(-1)^{i+j}
\ee
and

\be
\label{64}
\omega\tau  = \omega T (1 - \frac{\varepsilon}{2}\varepsilon_{ij}) -
\frac{\omega}{\omega_{0}}
\left(\varepsilon\varepsilon_{ij}(-1)^{i+j}\omega_{0}t +
\alpha_{ij}\right)
\ee
For small time $t$, we can write approximately:

\be
\label{65}
\cos(\omega_{0}\tau + \alpha_{ij})
\approx \cos\omega_{0} T(1 -
\frac{\varepsilon}{2}\varepsilon_{ij})
\ee
and from eq. {64}

\be
\label{66}
d\tau = dT(1 - \frac{\varepsilon}{2}\varepsilon_{ij})
\ee

So, in case of the excentric circles
the formula (64) can be obtained from nonperturbative formula (20)
only by transformation

\be
\label{67}
T \quad \longrightarrow \quad T(1 - \frac{\varepsilon}{2}\varepsilon_{ij}) ;
\quad \alpha_{ij} \quad\longrightarrow \quad
\left(\varepsilon\varepsilon_{ij}(-1)^{i+j}\omega_{0}t + \alpha_{ij}\right)
= \tilde {\alpha}_{ij}
\ee
excepting specific term involving sine functions.

Then, instead of formula (20) we get:

$$P(\omega,t) =
-\frac{\omega}{4\pi^2}\frac {e^{2}}{2R}
\*\frac{\mu}{n^2}\*\int_{-\infty}^{\infty} dT \sum_{i,j=1}^{2}(-1)^{i+j}
\times $$

\be
\label{68}
\cos(\omega T - \frac {\omega}{\omega_{0}}\tilde{\alpha}_{ij})
\left[1 - \frac {c^{2}}{n^{2}}v^{2}
\cos(\omega_{0} T)
\right]
\left\{\frac{\sin\left[\frac {2Rn\omega}{c}\sin
\left(\frac {\omega_{0}{\tilde T}}{2}\right)\right]}
{\sin\left(\frac {\omega_{0}{\tilde T}}{2}\right)}\right\}.
\ee
where $\tilde T = T(1 - \frac {\varepsilon}{2}\varepsilon_{ij})$.
We see that only $\tilde T$  and the $\alpha$ term
are the new modification of the original formula  (20).

However, because $\varepsilon$ term in the sine functions
is of very small influence on the behaviour of the total  function for
finite time $t$, we can neglect it and write approximatively:

$$P(\omega,t) =
-\frac{\omega}{4\pi^2}\frac {e^{2}}{2R}
\*\frac{\mu}{n^2}\*\int_{-\infty}^{\infty} dT \sum_{i,j=1}^{2}(-1)^{i+j}
\times $$

\be
\label{69}
\cos(\omega T - \frac {\omega}{\omega_{0}}\tilde{\alpha}_{ij})
\left[1 - \frac {c^{2}}{n^{2}}v^{2}\cos(\omega_{0} T)
\right]
\left\{\frac{\sin\left[\frac {2Rn\omega}{c}\sin
\left(\frac {\omega_{0}T}{2}\right)\right]}
{\sin\left(\frac {\omega_{0}T}{2}\right)}\right\}.
\ee

So, we se that only difference with the original radiation formula is in
variable  ${\tilde \alpha}_{ij}$.

It means that instead of sum (44) we have the following sum:

\be
\label{70}
\sum_{i,j = 1}^{2}(-1)^{i+j}\cos\frac {\omega}{\omega_{0}}
{\tilde\alpha_{ij}} =
2(1-\cos 2\omega t \cos\varepsilon\omega t)
\ee

It means that the one electron radiation formula is not
modulated by the formula
(44) but by the formula (70) and the final formula of for the power spectrum
is as follows:

\be
\label{71}
P_{l}(\omega,t) =
2(1-\cos 2\omega t \cos\varepsilon\omega t)
P_{l(electron)}\left(\omega\right),
\ee

For $\varepsilon \to 0$, we get the original formula (48).

\section{Discussion}

We have derived in this article the power spectrum formula
of the synchrotron radiation generated by the
electron and positron moving at the opposite angular
velocities in homogenous magnetic field. This article is an extended version
of the previous article [9].
We have used the Schwinger version of quantum field theory, for its simplicity.
It is suprising that the spectrum
depends periodically on time which means
that the system composed from electron, positron and magnetic
field behaves as a pulsar. While such pulsar can be represented by a
terrestrial experimental arrangement it is possible to consider also the
cosmological existence in some modified form.

To our knowledge, our result is not involved in the
classical monographies on the electromagnetic theory and at the same time
it was not still studied by the
accelerator experts investigating the synchrotron radiation of bunches.
This effect was not described in textbooks on
classical electromagnetic field and on the synchrotron radiation.
We hope that sooner or later this effect will be verified
by the accelerator physicists.

The radiative corrections
obviously influence the synergic spectrum of photons [2,7]. However,
the goal of this article is restricted only to the simple processes.

The particle laboratory  LEP in CERN  used instead of
single electron and positron the
bunches with 10$^{10}$ electrons or positrons  in one bunch of
volume 300$ \mu$m $\times$ 40$ \mu$m $\times$ 0.01 m.
So, in some approximation we can replace the charge of electron
and positron by the charges {Q} and {-Q} of both bunches in order to
get the realistic intensity of photons. Nevertheless the synergic character of
the radiation of two bunches moving at the opposite
direction in a magnetic field is conserved. The more exact description can
be obtained in case we consider the internal structure
of both bunches. But this is not goal of our article.\\

\end{document}